\documentstyle[12pt,psfig,rotating]{article}        
\newcommand{\chip}{\mbox{$\chi^+_1$}}  
\newcommand{\chiz}{\mbox{$\chi^0_1$}}  
\newcommand{\selec}{\mbox{$\tilde{e}$}}  
\newcommand{\squark}{\mbox{$\tilde{q}$}}  
\newcommand{\susyup}{\mbox{$\tilde{u}_L$}}  
\newcommand{\susytop}{\mbox{$\tilde{t}_1$}}  
\newcommand{\photino}{\mbox{$\tilde{\gamma}$}}  
\newcommand{\zino}{\mbox{$\widetilde{Z}$}}          
\newcommand{\higgsino}{\mbox{$\tilde{H}$}}      
\newcommand{\ra}{\mbox{$\rightarrow$}} 
\newcommand{\yukojk}{\mbox{$\lambda^\prime_{1jk}$}}  
\newcommand{\yukooo}{\mbox{$\lambda^\prime_{111}$}}  
\newcommand{\lumi}{\mbox{$\int\!{\cal L}dt$}} 
\newcommand{\pbarn}{\mbox{${\rm pb^{-1}}$}} 
\begin{document}
\title{Search for new physics at HERA\thanks{\bf OUNP-97-11}}
\author{Valerie A. Noyes \\
Particle and Nuclear Physics Laboratory\\
University of Oxford, Oxford OX1 3RH, UK}
\date{}
\maketitle

\begin{abstract}
Recent results on the searches for new physics in $ep$ collisions at HERA 
using the ZEUS and H1 detectors are presented. No evidence for 
excited fermions or supersymmetric particles within the context of the 
Minimal Supersymmetric Standard Model nor R-parity violation has been  found
using an integrated luminosity of up to 20~\pbarn. New limits have therefore been
established.
\vspace{1cm}\\
{\it Talk given at the Hadron Collider Physics 
conference, Stony Brook, June 1997, representing the ZEUS and H1 
collaborations}
\end{abstract}

\section{Introduction}
Due to the high centre of mass energy 
$\sqrt{s}=300~{\rm GeV}$ and presence of a lepton and quark in the initial 
state, searches for physics beyond the standard model (SM) in $ep$ 
collisions at  
HERA complement those made at LEP and the Tevatron. Indeed, much interest has
focused on the recent results published by the ZEUS and H1 
experiments~\cite{highx} in which an excess of events above standard model 
predictions was reported in the previously unexplored region of high $Q^2$.
No attempt will be made in this review, however, to discuss this
excess within the context of new physics. Further discussion and the latest
results in the high $Q^2$ region can be found in the conference 
proceedings~\cite{stony}.

Preliminary results are presented on the search for excited fermions
and selectron and squark production within the framework of 
R-parity conserving supersymmetric models using a total integrated luminosity
of 9.3~\pbarn\ and 20~\pbarn\ respectively. Results on the search for R-parity
violating squarks are also discussed taking into account all possible R-parity 
violating and gauge decays and using an integrated luminosity of 2.83~\pbarn.

\section{Search for excited fermions}
\begin{figure}[tbh]
\centerline{\psfig{file=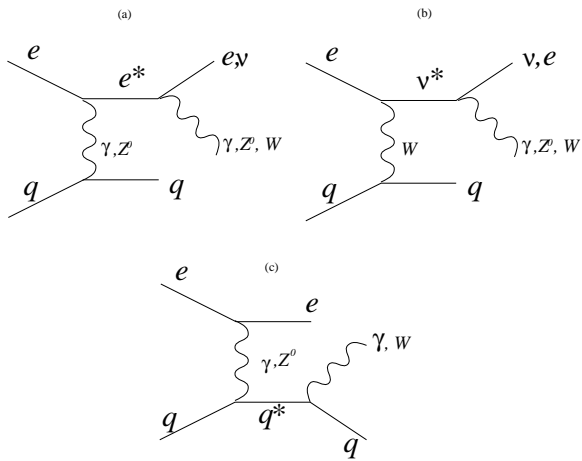,height=6cm}}
\caption{\label{fig:fstar_feyn} {Feynman diagram for excited 
(a) electron, (b) neutrino and (c) quark production at HERA. 
Only the decay modes considered in this search are shown.}}
\end{figure}

The existence of excited leptons or quarks would provide compelling evidence 
for a new layer of fermion structure predicted within theories of 
compositeness. At HERA single excited electrons, quarks and neutrinos can 
in principle be produced with masses up to the kinematic limit of 300 GeV via
the processes $ep \rightarrow f^* X, f^*\rightarrow fV$ shown in 
figure~\ref{fig:fstar_feyn}. The resonance is searched for via its 
electroweak decay to a light fermion, $f$, and a vector boson, $V$\@.

\begin{figure}[htb]
\centerline{\psfig{file=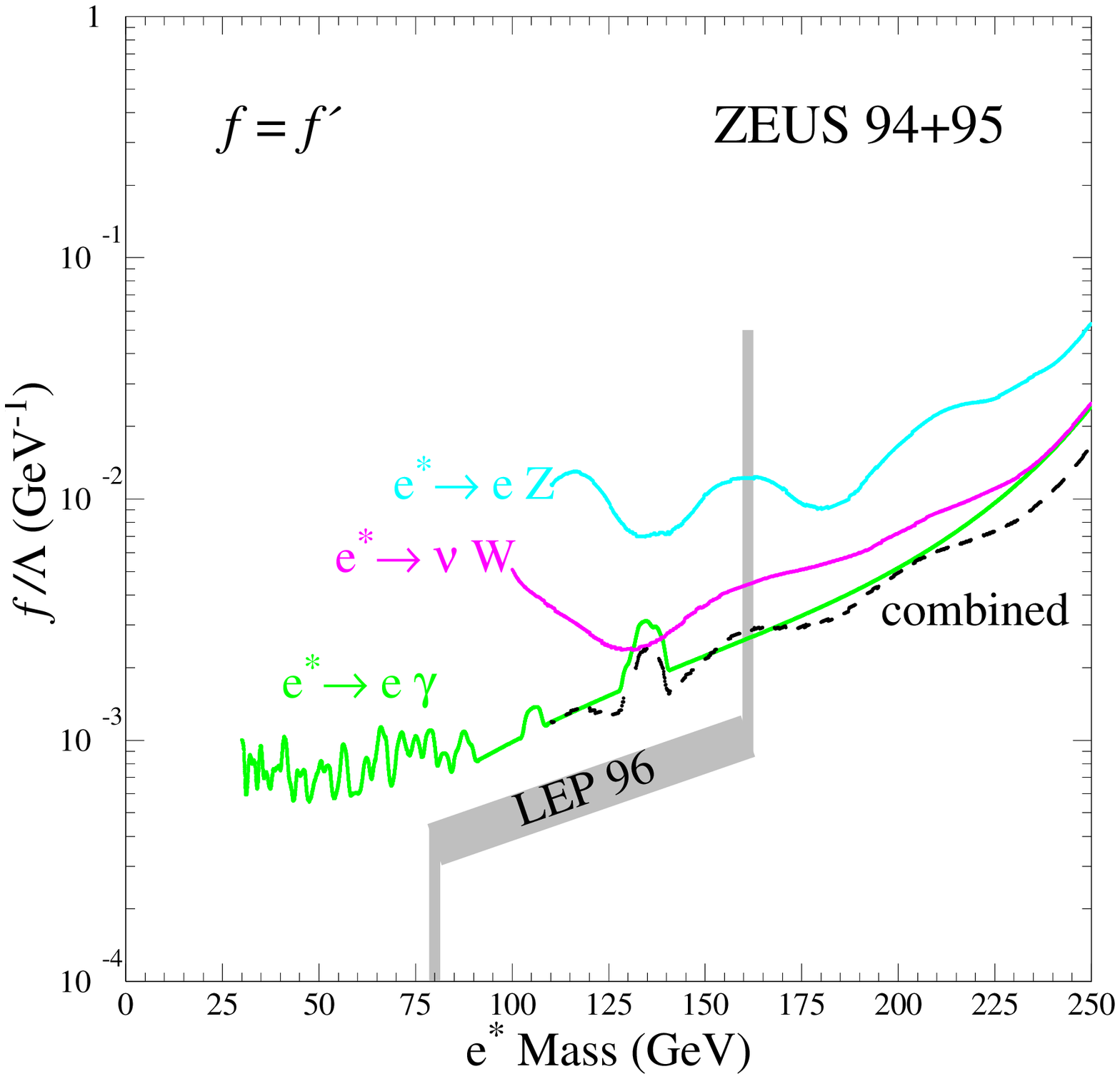,height=8cm}\psfig{file=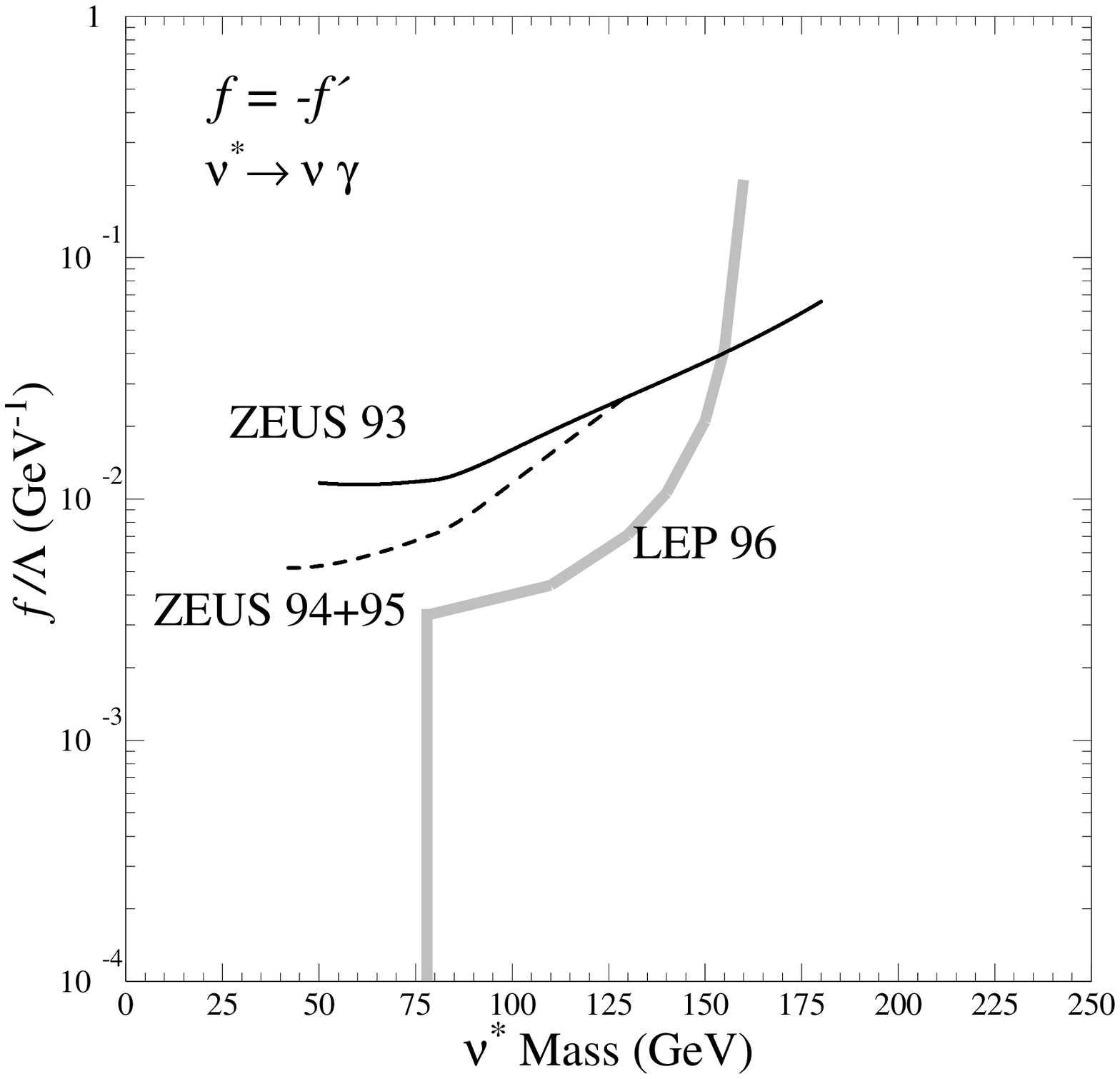,height=8cm}}
\centerline{\psfig{file=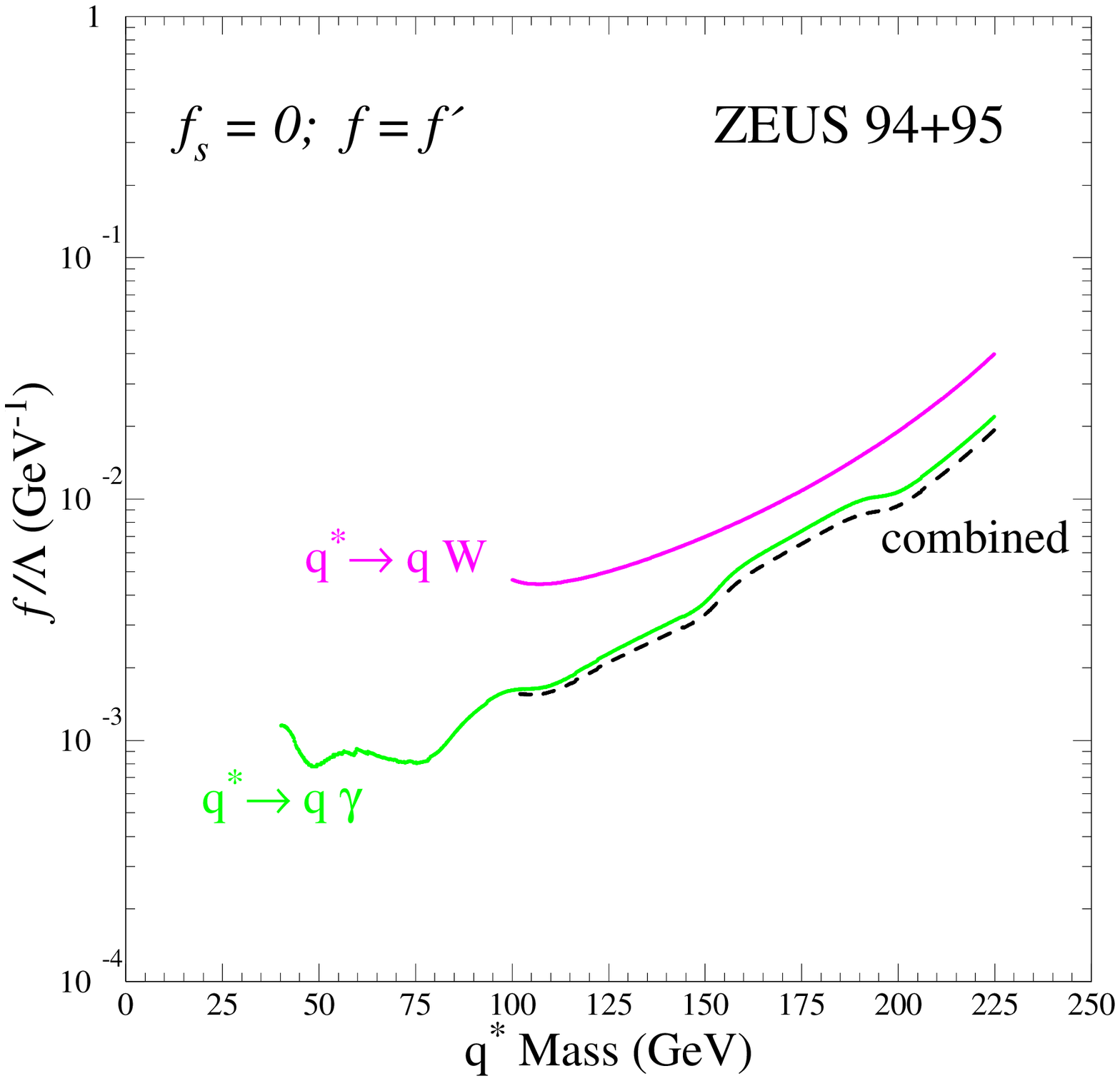,height=8cm}}
\caption{\label{fig:fstar} Limits on the characteristic coupling $f/ \Lambda$
as a function of the excited fermion mass for $e^*, \nu^*$ and $q^*$. The 
regions above the curves are excluded at 95\% CL. Also shown are exclusion 
limits from LEP with centre-of-mass energies up to 161 GeV and
the relationship between the couplings under which the limits were derived.}
\end{figure}

Direct searches for excited leptons and quarks have also been carried out at
LEP and the Tevatron. In $e^+e^-$ collisions, excited leptons may
be pair or singly produced up to $\sqrt{s}$. At 95\% confidence level (CL),
any excited lepton $(e^*, \mu^*, \tau^*, \nu^*$) with mass below approximately
80 GeV has been ruled out~\cite{lep_fstar}. This limit can be extended when 
the excited fermion is produced singly, however, becomes dependent on the 
coupling
parameters at the $ff^*V$ vertex as shown in figure~\ref{fig:fstar}. 
Experiments at the Tevatron have set stringent limits on the production of 
$q^*$ in $p\bar{p}$ collisions at $\sqrt{s}=1800~{\rm GeV}$, excluding
$q^*$ produced via the strong coupling~\cite{tev_fstar}  
with masses between 80 and 760~GeV, except for a window between 570 and 
580~GeV. HERA therefore extends the mass range over which $e^*$ and $\nu^*$ 
can be searched and offers unique sensitivity for excited quarks produced via
electroweak mechanisms.

In describing excited fermion production and decay it has become 
conventional to use the phenomenological model of Hagiwara et 
al~\cite{hagiwara}, extended  to include excited quark 
production by Baur et al~\cite{baur}.
The effective Lagrangian describing  magnetic transitions from spin
$\frac{1}{2}$ excited fermions $f^*$ to ordinary fermions $f$
has the form

\begin{equation}
L_{ff^*} = \frac{1}{\Lambda} \bar{f^*} \sigma^{\mu \nu}
\left[ gf \frac{\tau}{2}W_{\mu \nu} + g^\prime f^\prime \frac{Y}{2}B_{\mu \nu} + g_s f_s \frac{\lambda}{2}G_{\mu \nu} \right] f + h.c.
\end{equation}
where $\Lambda$ corresponds to the compositeness scale and the constants 
$f$, $f^\prime$ and $f_s$ describe the effective changes from the standard
model coupling constants $g$, $g^\prime$ and $g_s$. It is conventional to 
relate these unknown constants according to the chosen decay mode, thereby
reducing the cross section dependence to a single parameter, $f/\Lambda$.
For excited electron and excited quark searches 
$f=f^\prime$ and $f_s=0$ is chosen, exploring excited quark production via 
electroweak mechanisms. The radiative decay of excited neutrinos 
($\nu_e^* \rightarrow \nu_e \gamma$) is  only allowed when 
$f \neq f^\prime$ and in this case $f=-f^\prime$ and $f_s= 0$ is chosen. 

The ZEUS search for excited fermions in positron-proton collisions uses 
an integrated luminosity of 9.4~\pbarn. The $f^*$ decay modes which have 
been studied are shown in figure~\ref{fig:fstar_feyn} and the subsequent 
decay channels of the heavy boson are listed in table~\ref{table:fstar}.
A detailed description of the analysis is given in~\cite{zeus_fstar}.

\begin{table}[htb]
\begin{center}
\begin{tabular}{|c| c | c | c |} \hline
$f^*$ & \multicolumn{3}{|c|}{Final states} \\ \hline
$e^*$ & $e\gamma$ & $eZ^0$                    & $\nu W$ \\
      &           & $\rightarrow e q\bar{q}$  & $\rightarrow \nu q \bar{q}$ \\
      &           & $\rightarrow e\nu\bar{\nu}$ & $\rightarrow \nu e\nu$    \\\hline
$\nu^*$&$\nu\gamma$  & $\nu Z^0$                    & $e W$ \\
      &              & $\rightarrow \nu q \bar{q}$  &$\rightarrow e q\bar{q}$ \\
      &              &                              & $\rightarrow  ee\nu$ \\ \hline
$q^*$ & $q\gamma$ & $ qW$                        & \\
      &           & $\rightarrow qe \nu$          & \\ \hline
\end{tabular}
\caption{\label{table:fstar} The decay modes and subsequent final states 
considered in this search for excited fermions.} 
\end{center}
\end{table}
     
\begin{figure}[htb]
\centerline{\psfig{file=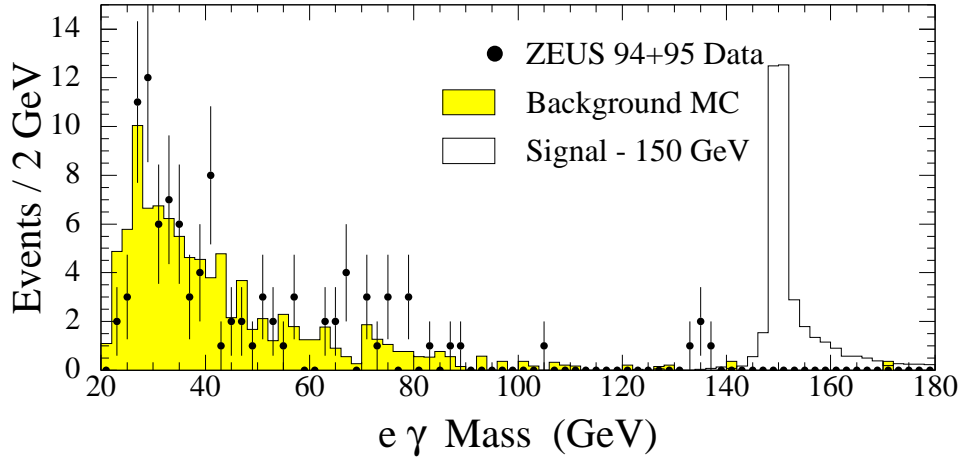,height=8cm}}
\caption{\label{fig:estarmass}Distribution of the $e\gamma$ invariant mass
for $e^* \rightarrow e\gamma$ candidates. The solid points show the ZEUS
data from 1994 and 1995 while the shaded histogram represents the expected
background. The unshaded histogram shows the expected distribution for
a 150 GeV excited electron.}
\end{figure}

Figure~\ref{fig:estarmass} shows the mass of the reconstructed $e\gamma$
final state in events with two isolated, high $E_T$ electromagnetic 
clusters. The data are compared to the predicted Monte Carlo background, 
comprising mainly QED Compton and neutral current DIS events. The unshaded
histogram shows the expected lineshape for an excited electron 
of mass 150 GeV. The cluster of four events at masses around 135 GeV
have features consistent with background expectations.

In the absence of a positive signal for excited leptons or quarks in any of 
the eight decay modes considered, upper limits 
on the characteristic 
coupling $f/\Lambda$ as a function of the excited fermion mass
were determined  at 95\% confidence level. Figure~\ref{fig:fstar}
shows the limits for $e^*$, $\nu^*$ and $q^*$ production derived 
from each of the final states considered and from their combination
using the assumptions about the couplings  discussed previously.  In each case 
the most stringent limits are derived from the $f^*\ra f\gamma$ channel and 
dominate  the combined limit from all three decay modes. The limits on
$e^*$ and $\nu^*$ production from the LEP experiments are also shown 
and demonstrate that HERA extends the limits for $f^*$ 
production well beyond the $\sim 170$ GeV limits currently 
achieved at LEP.

Also shown on the $\nu_e^*$ plot is the limit from the previous search by
ZEUS in $e^-p$ collisions with a much lower integrated luminosity (0.55~\pbarn).
This previous limit is more stringent for excited neutrino masses greater than
130 GeV because the $\nu_e^*$ production cross section in
$e^+p$ collisions is heavily suppressed relative to $e^-p$ collisions, 
out-weighing the increase in integrated luminosity taken with positrons.
High luminosity running with an $e^-$ beam planned for the 1998 run-period
will substantially improve this limit.

Exclusion limits on the excited fermion masses can be determined by further 
assuming that $f/\Lambda = 1/M_{f^*}$. For this case, at 95\% CL,
excited electrons are ruled out in the mass range between 30 and 200 GeV 
using the combined limit from all three decay modes. Excited neutrinos with 
masses in the range
40 to 96 GeV are excluded while excited quarks with electroweak-only couplings 
are excluded over the range 40 to 196 GeV. 

\section{Search for selectrons and squarks}
Supersymmetry (SUSY) is presently considered to be a promising candidate for 
a theory beyond the standard model. While there is strong theoretical 
motivation for its existence, there is currently no experimental 
evidence to support its prediction that each SM particle has a supersymmetric
partner differing by a half-unit of spin.

Within the minimal supersymmetric standard model (MSSM) it is 
assumed that R-parity is conserved. This implies that SUSY particles
can only be produced in pairs and that the lightest SUSY particle (LSP),
taken here to be the lightest neutralino \chiz\ is stable and weakly
interacting thereby escaping detection.

Existing limits from the LEP experiments at $\sqrt{s} =161~{\rm GeV}$
exclude selectrons with mass less than $\sim$~80~GeV~\cite{lep_select} and 
spartners to the light quarks with masses less that 
45 GeV (LEP1)~\cite{lep_squarks}. 
Although the experiments at the
Tevatron set strong limits on squark masses, these are dependent on
the gluino mass~\cite{tev_squark} and can only be related to the HERA results 
assuming additional relations motivated by Grand Unified Theories (GUT). 

\begin{figure}[htb]
\centerline{\psfig{file=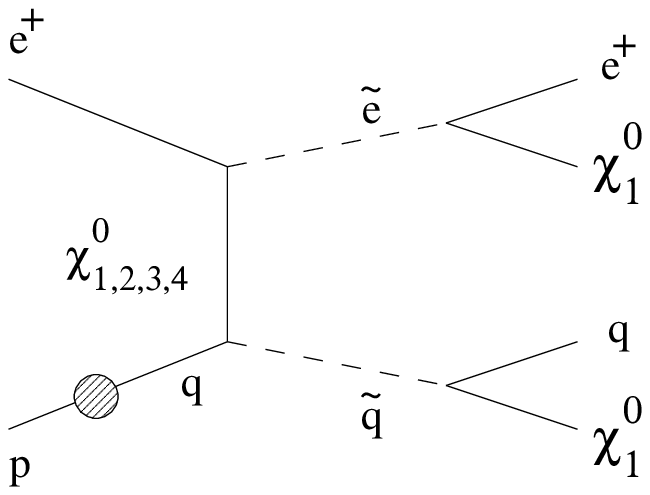,height=5cm}}
\caption{\label{fig:mssm_feyn} Selectron-squark production via neutralino
exchange and the following decays to the LSP and SM partner.}
\end{figure}

At HERA the dominant MSSM process is the production of a selectron and
squark via $t$ channel neutralino exchange $ep\rightarrow \selec\squark X $ 
as shown in figure~\ref{fig:mssm_feyn}. Although the \selec\ and \squark\ 
can decay into any lighter gaugino and their SM partners, the cleanest
experimental signature involves both the $\selec \rightarrow e \chiz$ and
$\squark \rightarrow q \chiz$ decays and is the one considered here. 
The final state topology consists therefore of 
an electron which is acoplanar to the hadronic system and missing momentum.

Limits from a earlier search using the H1 detector and an integrated 
luminosity of 6.4~\pbarn\ have been published in~\cite{h1selectron}. The current 
search by ZEUS uses an integrated luminosity of 20~\pbarn.
No signal was observed and preliminary exclusion limits on 
\selec\squark\ production were derived at $95\%$ confidence level. The limits
are interpreted as exclusion regions in MSSM parameter space. The
quantities used are $M_2$ and $M_1$, the mass parameters
for the $SU_2$ amd $U_1$ gauginos, the higgsino mass parameter $\mu$,
and $\tan \beta \equiv v_2/v_1$, the ratio of the 
two Higgs doublet vacuum expectation values.

\begin{figure}[tbh]
\centerline{\psfig{file=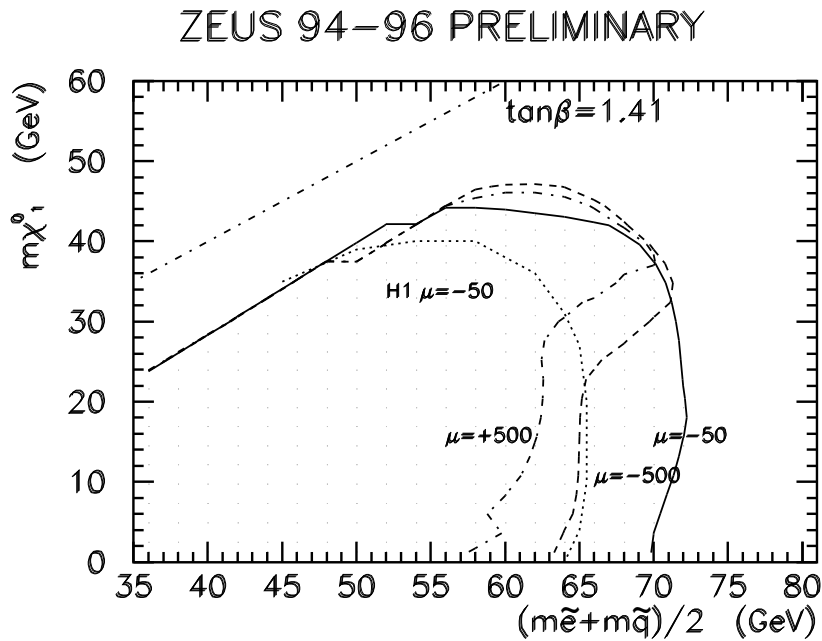,height=10cm}}
\caption{\label{fig:m0mq} Regions excluded at 95\% CL in the plane defined by
the lightest neutralino mass and half the sum of the selectron and 
squark masses for $\mu$=-500 (dashed), -50 (full), +500 (dash-dotted line)
and $\tan \beta=1.41$.  The limit from an earlier search by 
H1
is also shown for $\mu=-50$~GeV (dotted line).}
\end{figure}

Figure~\ref{fig:m0mq} shows the excluded region in the plane defined by the
mass of the lightest neutralino and half the sum of the selectron and
squark masses for $\tan \beta=1.41$ and different values of $\mu$.
For small values of $\mu$ ($\mu=-500~{\rm GeV}$) and low neutralino masses,
competing decays of the $\selec$ and $\squark$ to gauginos other 
than the \chiz\ lead to reduced limits for $(M_{\selec}+M_{\squark})/2$.
In the region $\mu=-50~{\rm GeV}$, the branching ratios $B(\selec\ra e \chiz)$
and $B(\squark \ra q \chiz)$ are close to one because the other gauginos 
are heavier than $M_{\selec}$ or $M_{\squark}$. At large $M_{\chiz}$ the
mass difference between the \selec, \squark\ and \chiz\ becomes small leading
to a reduction in cross section together with a drop in detection efficiency
limiting the upper exclusion bound.
The excluded mass range extends to 72~GeV for $(M_{\selec}+M_{\squark})/2$
and 47~GeV for $M_{\chiz}$, extending the 
previous limits set by H1~\cite{h1selectron}  with lower luminosity.

\begin{figure}[tbh]
\centerline{\psfig{file=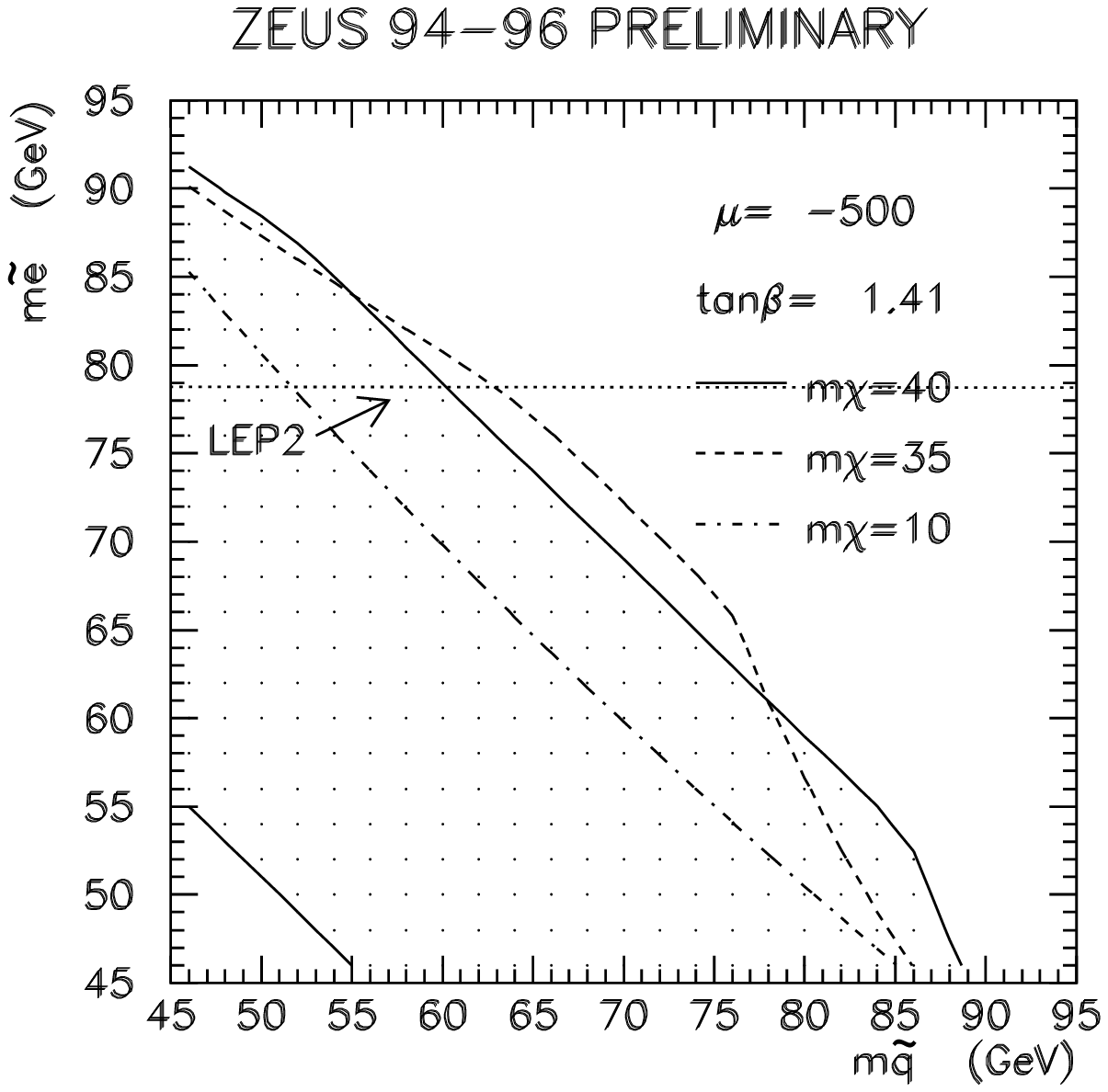,height=10cm}}
\caption{\label{fig:memq} Regions excluded at 95\% CL in the plane defined
by the selectron and squark mass for fixed values of neutralino mass
$M_{\chiz}$=10 (dash-dotted), 35 (dashed), 40 (full line)~GeV and 
for $\mu=-500$~GeV  and $\tan \beta=1.41$. Selectron mass limits
from LEP2 are also shown.}
\end{figure}

Figure~\ref{fig:memq} shows the excluded region in $M_{\selec}$
versus $M_{\squark}$ for different values of fixed \chiz\ mass at 
$\mu=-500~{\rm GeV}$.
The limits are approximately functions of the sum $M_{\selec}+M_{\squark}$
due to the cross section and efficiency depending almost essentially on this
sum. For $M_{\chiz}=40~{\rm GeV}$ squarks with masses between 
55~GeV and 89~GeV are excluded for $M_{\selec}=46~{\rm GeV}$ while
selectrons with mass between 55~GeV and 92~GeV are excluded for
$M_{\squark}=46~{\rm GeV}$. The selectron mass exclusion 
limits from LEP2~\cite{lep_select} are also indicated in 
figure~\ref{fig:memq} and show that
the new HERA limits exclude a further region at high selectron and low 
squark masses.

\begin{figure}[tbh]
\centerline{\psfig{file=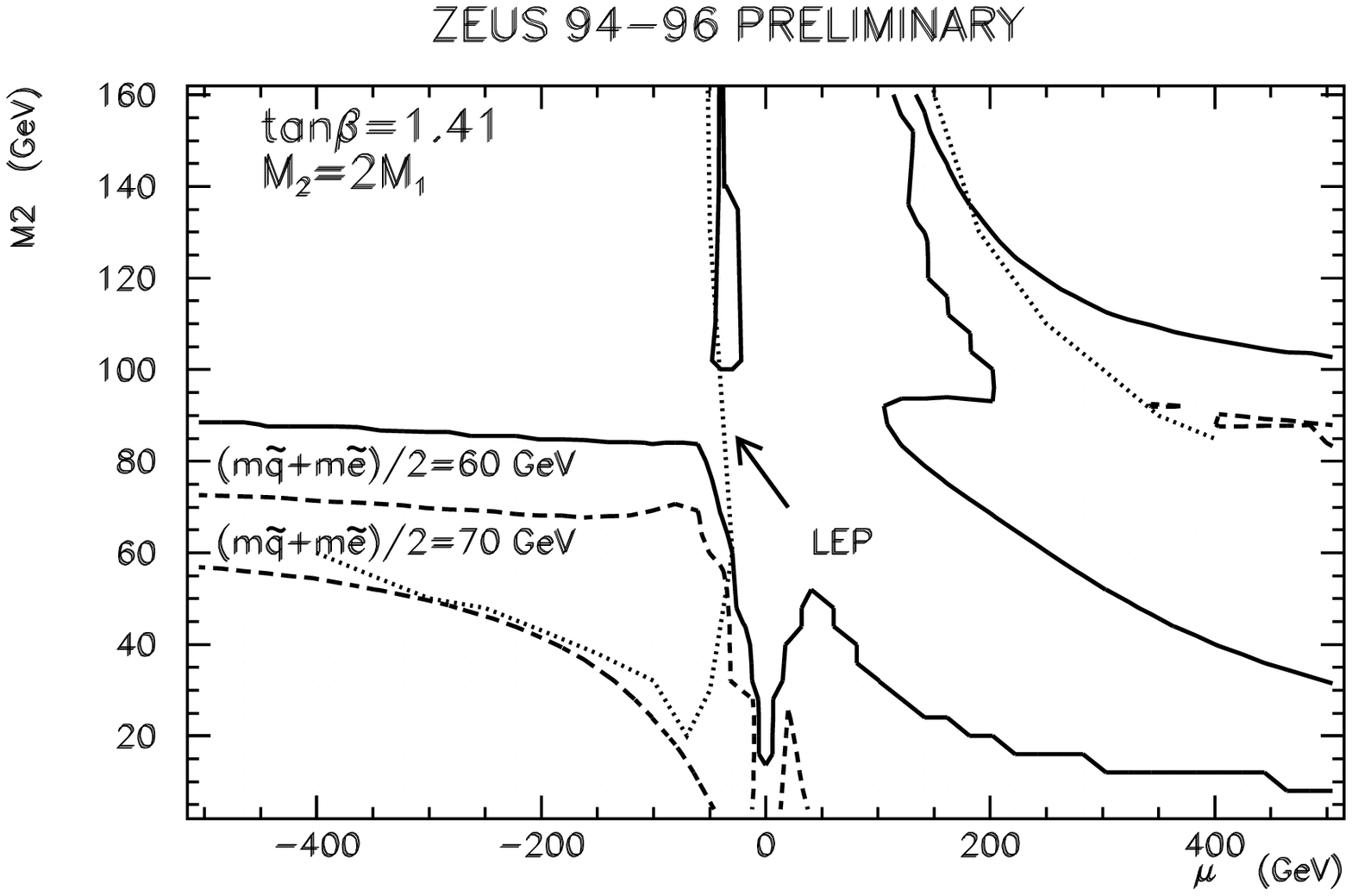,height=10cm}}
\caption{\label{fig:mum2} Region excluded at 95\% CL in the plane defined
by $M_2$ and  $\mu$ for $M_{\selec}=M_{\squark}$=~60 (full) and
70 (dashed line)~GeV for $\tan \beta=1.41$. The area below the dotted
line is excluded by LEP.}
\end{figure}

In figure~\ref{fig:mum2} the exclusion limits on $M_2$ versus $\mu$ are shown
for $\tan \beta=1.41$ and $M_2=2M_1$, a good approximatation for the only
SUSY GUT relation used here $M_1 = \frac{5}{3} M_2\tan^2 \theta_W$. For 
$\mu \ll 0$, the \chiz\ is dominated by its photino component leading to 
couplings to the selectron and squark that are electromagnetic in strength 
and allow for a 
sizable cross section. As $\mu \sim 0$ the \chiz\ becomes higgsino-like 
and the couplings and cross section become very small. The ZEUS limits are
compared to those from chargino and neutralino searches at LEP~\cite{aleph}
and extend considerably  the limits on $M_2$ at negative $\mu$.

\section{Search for R-parity violating squarks}
\begin{figure}[htb]
\centerline{\psfig{file=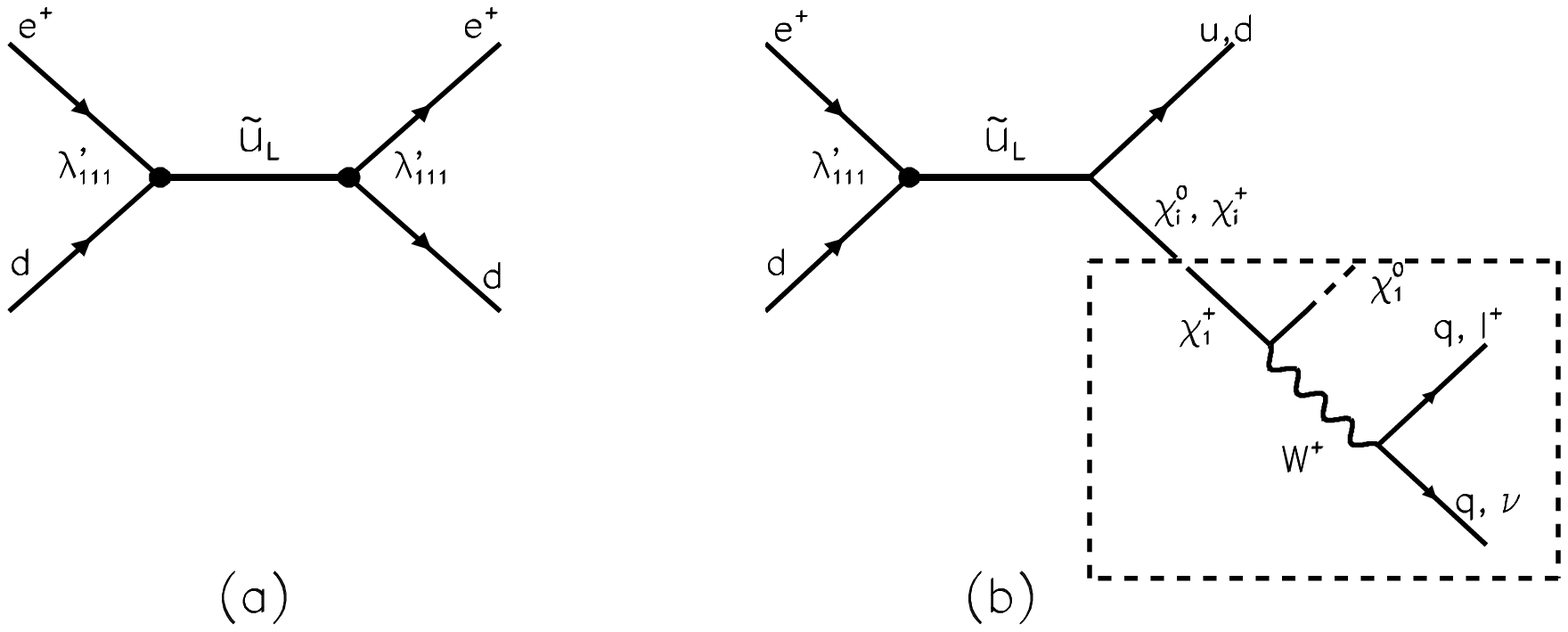,height=5cm}}
\centerline{\psfig{file=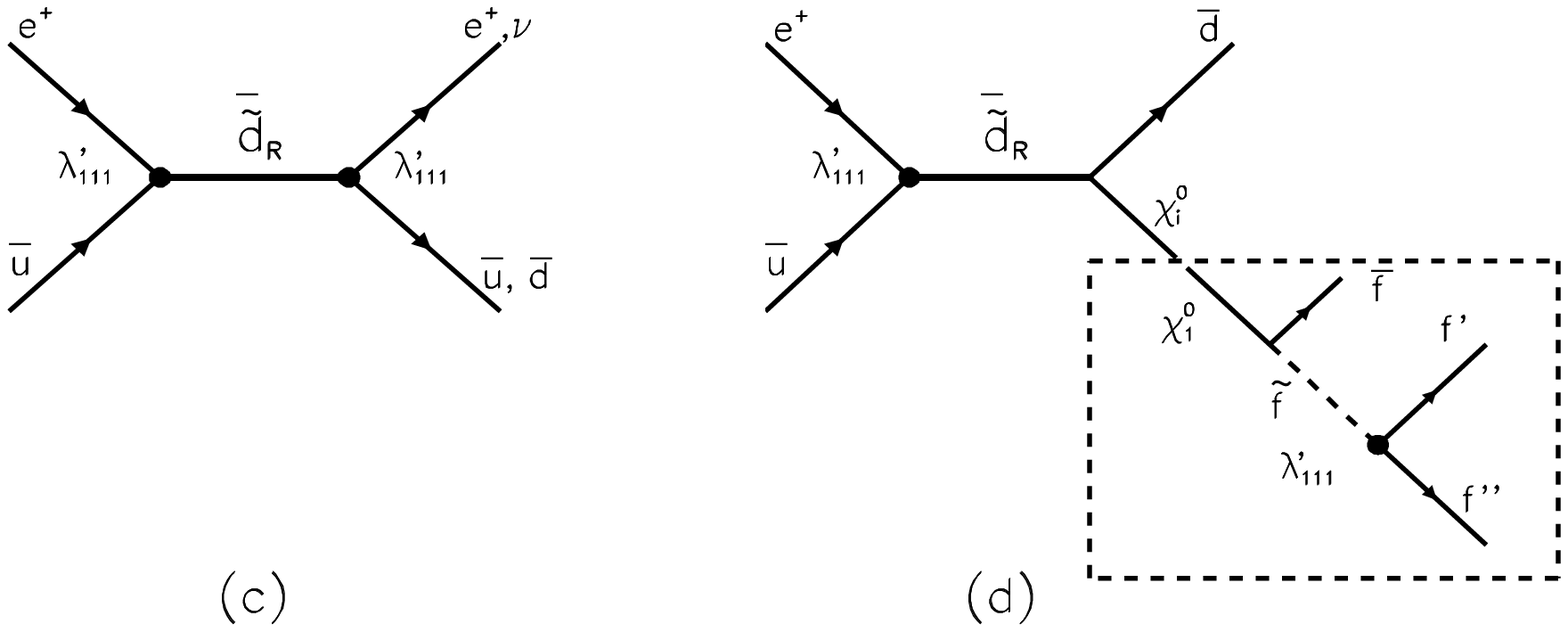,height=5cm}}
\caption{\label{fig:rparity_feyn} First generation R-parity violating 
squark production at HERA followed by (a), (c) R-parity violating decays and
(b) and (d) gauge decays. In (b) and (d) the chargino and neutralino 
undergo subsequent R-parity violating decays.}
\end{figure}

The search for R-parity violating squarks possessing a Yukawa coupling 
$\lambda^\prime$ to lepton-quark pairs is particularly promising at 
HERA. 
Such squarks can be singly produced as an $s$ channel resonance 
up to the kinematic limit of $\sqrt{s}=300 ~{\rm GeV}$ 
via positron-quark fusion. The squarks subsequently decay via their Yukawa 
coupling into fermions or via their gauge couplings into a quark and
neutralino or a chargino as shown in figures~\ref{fig:rparity_feyn}(a), (c) 
and (b), (d) respectively. The $\chi^0_i$ and $\chi^+_i$ are
mixed states of the supersymmetric partners to the SM gauge bosons and
neutral Higgses and are in general unstable. In contrast to the MSSM, 
this also holds for the LSP which can decay into a quark, antiquark and lepton.

In the case where both the squark production and decay occur through a Yukawa
\yukojk\ coupling \footnote{the $ijk$ indices correspond to to the generations 
of superfields  $L_i, Q_j$ and $\bar{D}_k$ containing the
left-handed lepton doublet and  quark doublet, and the right-handed quark 
singlet respectively} the final state 
signatures consist of a lepton and jet
and are indistinguishable from standard neutral and
charged current deep inelastic scattering on an event-by-event basis. The 
strategy then involves searching for a resonance at high mass ($M_{\squark} = 
\sqrt{xs}$) where $x$ is the Bjorken scaling variable. No attempt is
made here to interpret the recently
reported excess of events at high $x$ and $Q^2$ observed at the HERA 
experiments~\cite{highx} within the context of R-parity violating SUSY.

\begin{figure}[htb]
\centerline{\psfig{file=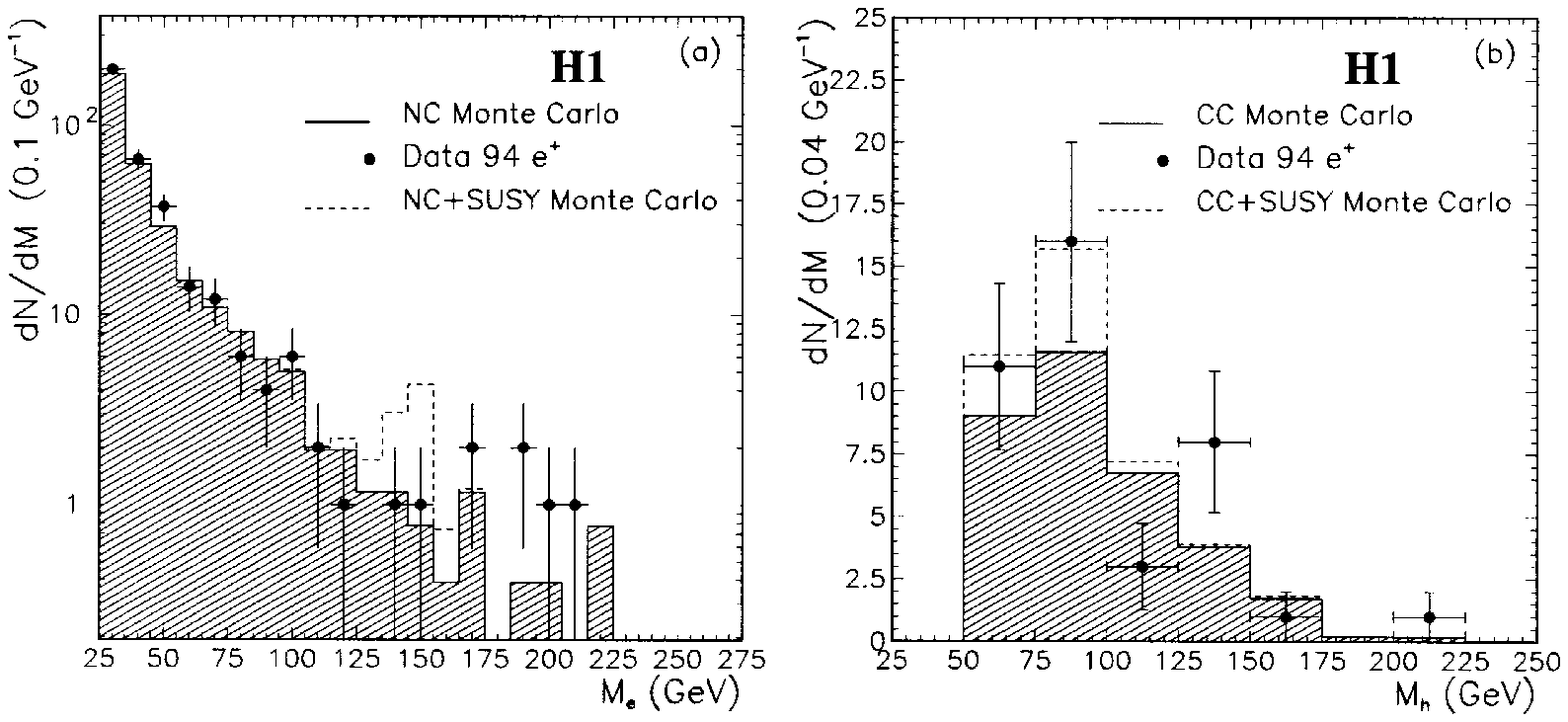,width=15cm}}
\caption{\label{fig:h1_squark} Reconstructed  mass distribution in the
(a) neutral and (b) charged current event samples using an integrated luminosity
of 2.8~\pbarn. The shaded histogram shows the background MC spectrum while
the dashed histogram indicates the mass distribution of a mixture of background
MC and a SUSY signal with $M_{\squark}=150~{\rm GeV}$ and 
$M_{\squark}=75~{\rm GeV}$ in (a) and (b) respectively.}
\end{figure}

Using an integrated luminosity of 2.8~\pbarn\ the squark mass is reconstructed 
using the $e^+$ energy and polar angle in the selected neutral current
sample and the hadronic final state in the charge current sample.
Figures~\ref{fig:h1_squark}(a) and (b) show the reconstructed mass for the 
neutral and charged current event samples respectively. The shaded histogram
indicates the background Monte Carlo while the dashed histogram compares
the mass distribution for a mixture of background and SUSY MC events for
$M_{\squark}=150~{\rm GeV}$ and $M_{\squark}=75~{\rm GeV}$ respectively.
No significant deviation from the number of expected events is observed when
including all sources of systematic error.

Squark decays via gauge couplings $\tilde{u}_L\rightarrow u\chi^0_i$ or
$d\chi^+_j$ and $\bar{\tilde{d}}_R\rightarrow \bar{d}\chi^0_i$ have also been
considered (figure~\ref{fig:rparity_feyn}~(b) and (d)). Furthermore, both 
gauge and R-parity violating decays of the  chargino have been taken into 
account 
together with the dependence of
the \chiz\ decay on its gaugino-higgsino composition governed by the 
fundamental SUSY parameters. In particular if the \chiz\ is dominated by
its photino component it will tend to decay 
$\chiz \rightarrow e^\pm q\bar{q}$, giving the possibility of a ``wrong sign''
lepton (with respect to the charge of the incoming lepton beam) in the 
final state.

\begin{table}[htb]
\begin{center}
\begin{tabular}{|l | c | l c l c l | c|} \hline
Topology &       \chiz\ nature         & \multicolumn{5}{|c|}{Decay process}                     & Signature \\ \hline 
\rule[-3mm]{0mm}{8mm}   a        & \photino, \zino            & $\squark$ & \ra & $ q \chiz$ & \ra & $q e^+\bar{q}q$ & High $p_T$ $e^+$ \\
         & \photino, \zino, \higgsino & $\susyup$ & \ra & $ d \chip$ & \ra & $de^+d\bar{d}$  &       +          \\
         & \photino, \zino            & $\susyup$ & \ra & $d \chip$  & \ra & $W^+\chiz $     & multiple jets     \\
         &                            &           &     &            & \ra & $q\bar{q}e^+\bar{q}q $          &                  \\\hline
\rule[-3mm]{0mm}{8mm}   b        & \photino, \zino            & $\squark$ & \ra & $q \chiz$  & \ra & $q e^-\bar{q}q$ & High $p_T$ $e^-$ \\
         & \photino, \zino            & $\susyup$ & \ra & $d \chip$  & \ra & $W^+\chiz $     &           +      \\
         &                            &           &     &            & \ra & $q\bar{q}e^-\bar{q}q $                 & multiple jets    \\\hline
\rule[-3mm]{0mm}{8mm}   c        & \photino, \zino            & $\squark$ & \ra & $q \chiz$  & \ra & $q \nu\bar{q}q$ &                 \\
         & \photino, \zino            & $\susyup$ & \ra & $d \chip$  & \ra & $W^+ \chiz $    & Missing $p_T$    \\
         &                            &           &     &            & \ra & $q \bar{q} \nu \bar{q} q$   &      +           \\
         & \photino, \zino, \higgsino & $\susyup$ & \ra & $d \chip$  & \ra & $d \nu u \bar{d}$& multiple jets    \\
         & \higgsino                  & $\susyup$ & \ra & $d \chip $ & \ra & $W^+ \chiz $    &                  \\
         &                            &           &     &            & \ra & $ q\bar{q}\chiz$                       &                  \\\hline
\rule[-3mm]{0mm}{8mm}   d        & \higgsino                  & $\susyup$ & \ra & $d\chip $  & \ra & $W^+\chiz$      & High $p_T$ $e^+$ or $\mu^+$ \\
         &                            &           &     &            & \ra & $l^+ \nu \chiz$                 & missing $p_T$ + 1 jet\\\hline
\rule[-3mm]{0mm}{8mm}   e        & \photino, \zino            & $\susyup$ & \ra & $d\chip $  & \ra & $W^+ \chiz$     & High $p_T$ $e$ + $e^+$ or $\mu^+$ \\
         &                            &           &     &            & \ra & $l^+ \nu e^\pm \bar{q} q$   & + missing $p_T$ \\
         &                            &           &     &            &     &                 & + multiple jets \\\hline
\rule[-3mm]{0mm}{8mm}   f        & \photino, \zino            & $\susyup$ & \ra & $d\chip $  & \ra & $W^+ \chiz$     & High $p_T$ $e^+$ or $\mu^+$ \\
         &                            &           &     &            & \ra & $l^+ \nu \nu \bar{q} q$         & + missing $p_T$ \\
         &                            &           &     &            &     &                 & + multiple jets \\ \hline
\end{tabular}
\caption{\label{table:susy_top} Squark decay channels in R-parity violating 
SUSY classified according to distinguishable event topologies.}
\end{center}
\end{table}

\begin{figure}
\centerline{\psfig{file=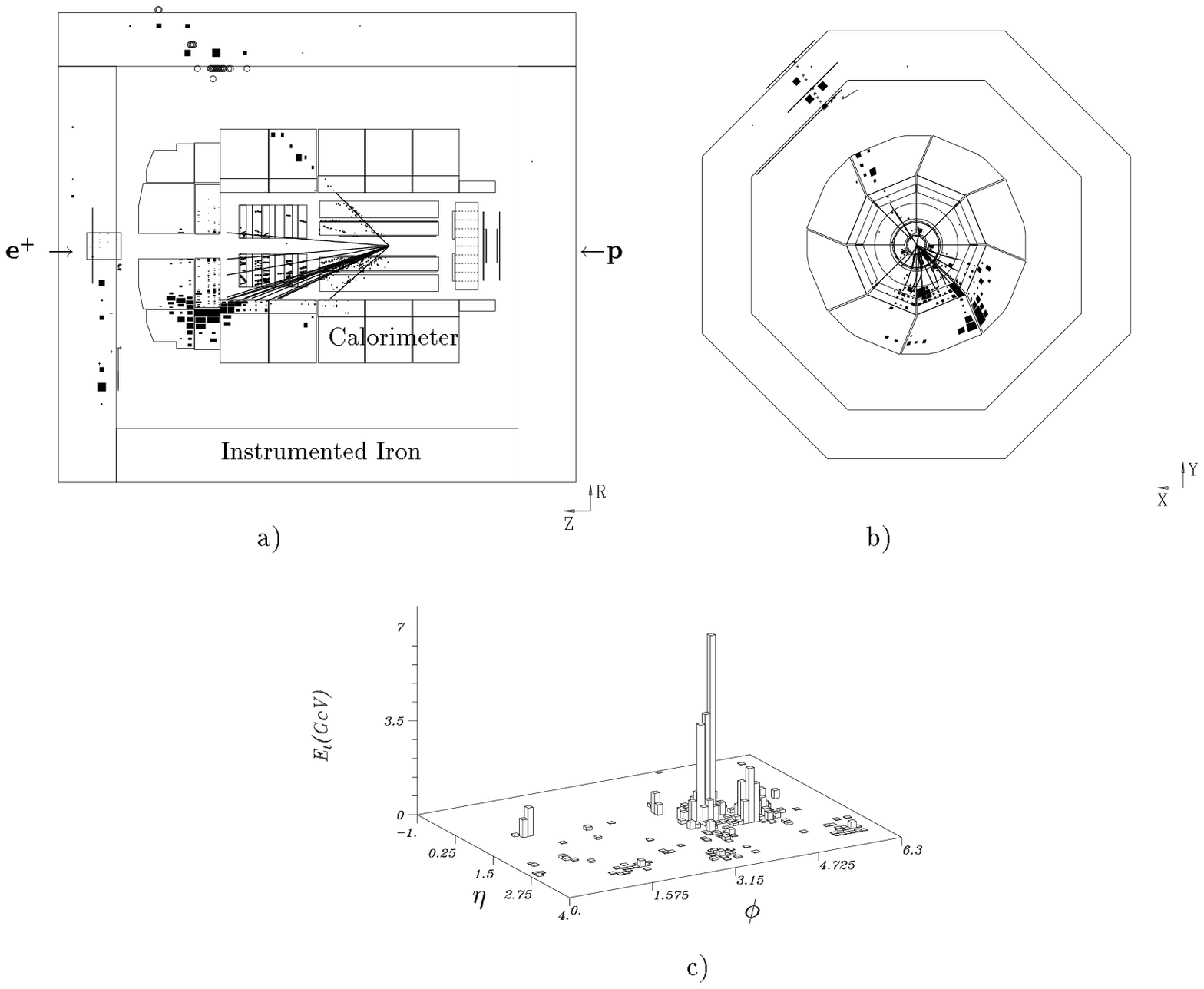,height=13cm}}
\caption{\label{fig:theevent} Display of the 
$e^+p\ra \mu+jet$ candidate found using the H1 detector.}
\end{figure}

\begin{figure}
\centerline{\psfig{file=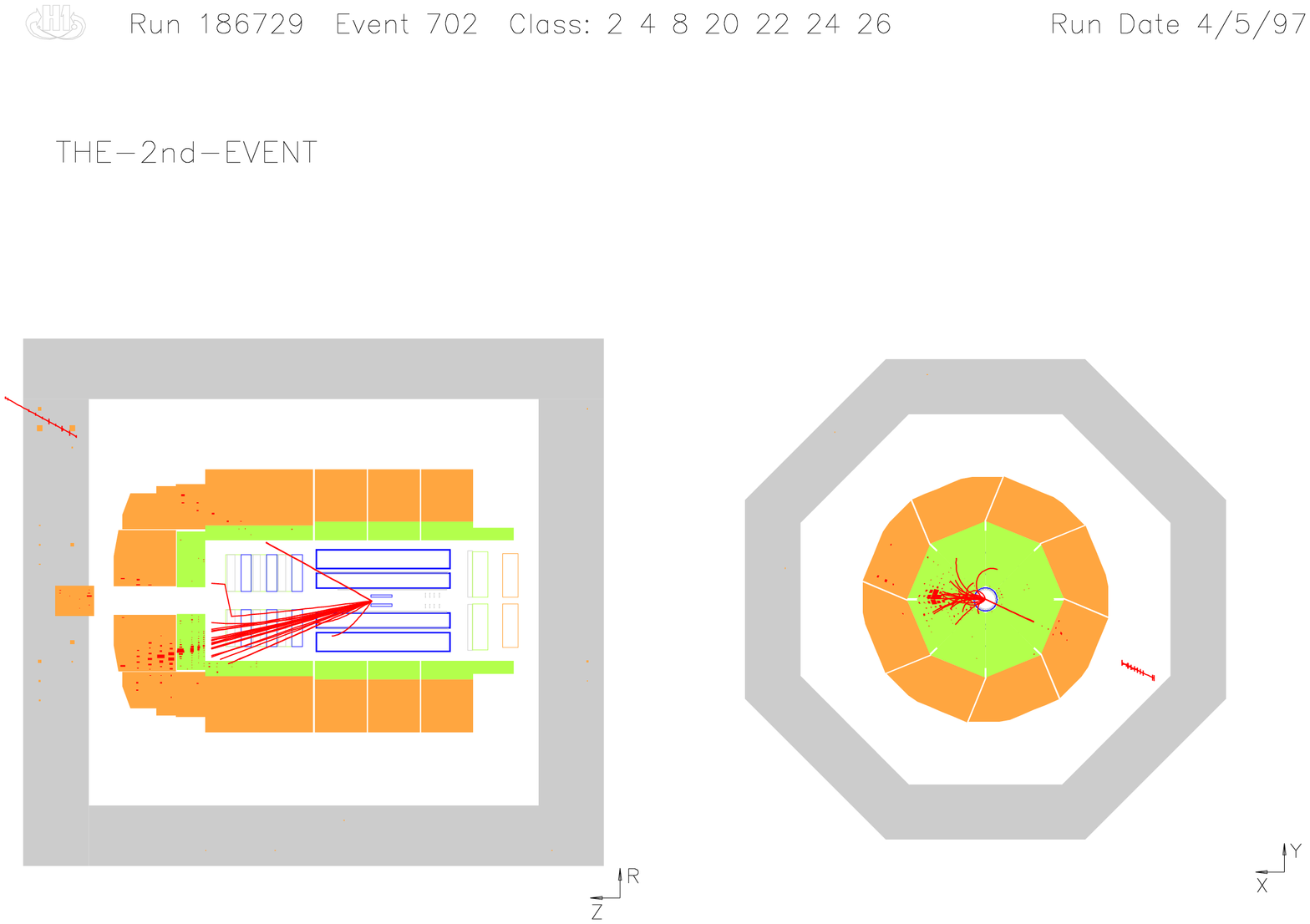,angle=90,height=13cm}}
\caption{\label{fig:the2ndevent} A second $e^+p\ra\mu + jet$ candidate 
recently found at H1.}
\end{figure}

Six distinguishable event topologies are listed in 
table~\ref{table:susy_top} and have been considered in this search, 
full details of which are given in~\cite{h1squark}.
No significant excess of events was observed in any channel, however, an 
interesting event with a final state $\mu$ and jet passed the selection 
cuts for topology d in 
table~\ref{table:susy_top} and is shown in figure~\ref{fig:theevent}. 
Detailed analysis of the event~\cite{h1event} shows the $\mu^+$ has a 
$p_T=23~{\rm GeV}$ and there is an overall missing $p_T=18.7~{\rm GeV}$, 
taking the 
$p_T$ of the $\mu^+$ into account. The dominant source of background is 
expected from single $W^+$ production followed by the decay 
$W^+ \ra \mu^+\nu X$. However the properties of this event are such that there 
is only a $\sim 3\%$ probability for such an interpretation and a negligible 
contribution from misidentified charged current DIS.  A second similar
candidate has been observed during 1997 data-taking and is shown in 
figure~\ref{fig:the2ndevent}, however more data will be required to
establish the origin of these events.

\begin{figure}[tbh]
\centerline{\psfig{file=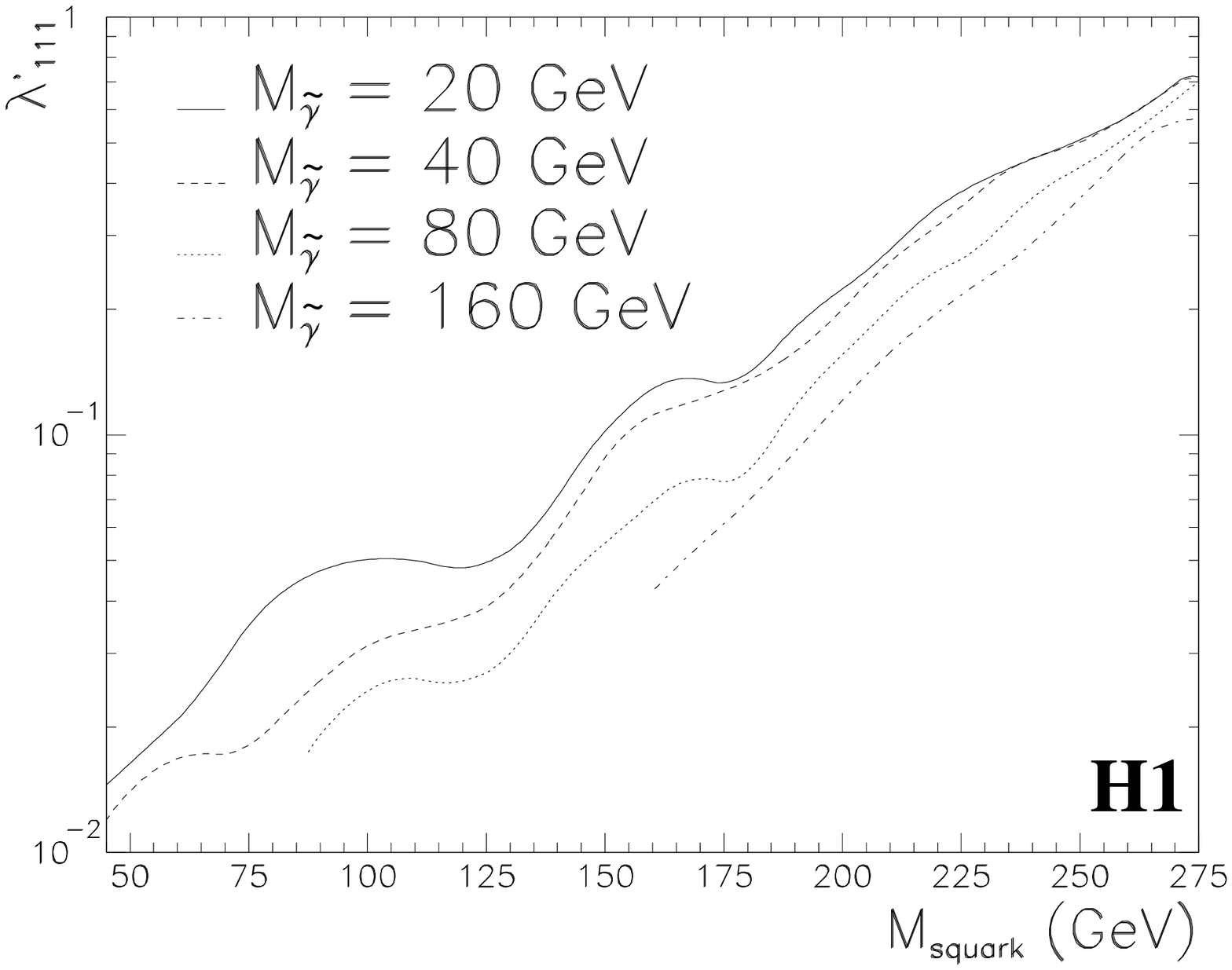,height=8cm}}
\caption{\label{fig:h1squark_limits} Exclusion limits for the  
squark coupling \yukooo\ as a function of squark mass for various fixed
photino masses and taking $\tan\beta=1$. The regions above the curves are
excluded at 95\% CL and combine R-parity violating and gauge decays of the
squarks.}
\end{figure}

In the absence of a significant deviation from SM expectations, 
95\% confidence level limits limits on the Yukawa coupling \yukooo\ are 
derived as a function of squark mass combining all contributing channels. 
Figure~\ref{fig:h1squark_limits} shows the
limits assuming that the lightest neutralino is a pure photino and for 
different photino masses. First generation squarks with masses up to 240~GeV 
are excluded at 95\% confidence level 
for coupling strengths $\lambda^{\prime 2}_{111}/4\pi > \alpha_{em}$, 
thereby extending
the limits inferred from the Tevatron di-lepton data~\cite{tev_dilepton}.

From the analysis of the \yukooo\ case, limits on the \yukojk\ couplings can 
be  deduced by folding in the proper parton densities. The results given in 
table~\ref{table:yukawa} are for $M_{\squark}=150~{\rm GeV}$ and 
$M_{\chiz}=80~{\rm GeV}$ for \photino-dominant and \zino-dominant \chiz.
No other direct limits for \yukojk\ where $j$ or $k \neq 1$ have been 
published.

\begin{table}[htb]
\begin{center}
\begin{tabular}{| l |l@{\ra}l| l| l|} \hline
$\lambda^\prime_{1jk}$  & \multicolumn{2}{|c|}{production} & \photino-like \chiz & \zino-like \chiz     \\
                        & \multicolumn{2}{|c|}{processes} &                      &                  \\ \hline 
\rule[-3mm]{0mm}{8mm}   $\lambda^\prime_{111}$  & $e^+ + \bar{u}$ & $\bar{\tilde{d}}_R$ & 0.056               & 0.048                \\
			& $e^+ + d$       & $\tilde{u}_L$       &                     &      \\\hline
\rule[-3mm]{0mm}{8mm}   $\lambda^\prime_{112}$  & $e^+ + \bar{u}$ & $\bar{\tilde{s}}_R$ & 0.14                & 0.12                 \\
                        & $e^++ s$        & $\tilde{u}_L$       &                     &      \\\hline
\rule[-3mm]{0mm}{8mm}   $\lambda^\prime_{113}$	& $e^++ \bar{u}$  & $\bar{\tilde{b}}_R$	& 0.18                & 0.15                 \\
                        & $e^++ b$        & $ \tilde{u}_L$      &                     &      \\\hline
\rule[-3mm]{0mm}{8mm}   $\lambda^\prime_{121}$  & $e^++ \bar{c}$  & $\bar{\tilde{d}}_R$ & 0.058               & 0.048                \\
                        & $e^++ d $       & $\tilde{c}_L$       &                     &      \\\hline
\rule[-3mm]{0mm}{8mm}   $\lambda^\prime_{122}$  & $e^++ \bar{c}$   & $\bar{\tilde{s}}_R$ & 0.19                & 0.16                 \\
                        & $e^++ s$         & $\tilde{c}_L$       &                     &   \\\hline
\rule[-3mm]{0mm}{8mm}   $\lambda^\prime_{123}$  & $e^++ \bar{c}$  & $\bar{\tilde{b}}_R$	& 0.30                & 0.26                 \\
                        & $e^+++ b$       & $\tilde{c}_L$       &                     & \\\hline 
\rule[-3mm]{0mm}{8mm}   $\lambda^\prime_{131}$  & $e^++\bar{t}$   & $\bar{\tilde{d}}_R$ & 0.06                & 0.05                 \\
			& $e^++ d$        & $\tilde{t}_L$	&                     & \\\hline 

\rule[-3mm]{0mm}{8mm}   $\lambda^\prime_{132}$  & $e^++\bar{t}$   & $\bar{\tilde{s}}_R$	& 0.22                & 0.19                 \\
                        & $e^++ s$        & $\tilde{t}_L$      &                     & \\\hline 
\rule[-3mm]{0mm}{8mm}   $\lambda^\prime_{133}$  & $e^++ \bar{t}$  & $\bar{\tilde{b}}_R$	& 0.55                & 0.48                 \\
               		& $e^++ b $       & $\tilde{t}_L$	&		      & \\\hline 
\end{tabular}
\caption{\label{table:yukawa} Upper 95\% CL exclusion limits on the couplings
\yukojk\ for $M_{\squark}=150~{\rm GeV}$ and $M_{\chiz}=80~{\rm GeV}$.}
\end{center}
\end{table}

\section{Search for stop}
\begin{figure}[htb]
\centerline{\psfig{file=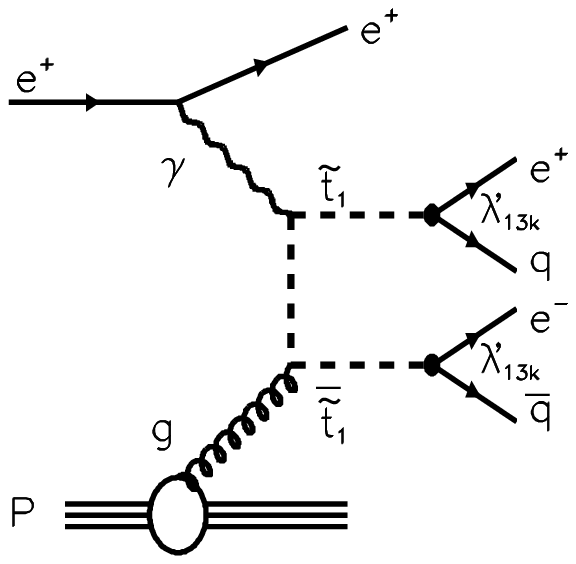,height=8cm}}
\caption{\label{fig:stop_feyn} Stop pair production via photon-gluon
fusion followed by the subsequent R-parity violating decay $\susytop \ra eq$.}
\end{figure}

The cases where $\lambda_{131}^\prime \neq 0$ or $\lambda_{131}^\prime \neq 0$
are of particular interest as they allow for the direct production of stop. 
A light stop mass eigenstate, \susytop, much lighter than the top quark or 
other squarks could exist.
The search for such a stop mass eigenstate is extended towards 
lower masses by considering $\tilde{t}_1\bar{\tilde{t}}_1$
pair production via photon-gluon fusion as shown in figure~\ref{fig:stop_feyn}.
Here the \susytop\ is assumed to be lighter than the top quark and the 
lightest chargino so that the decays into $t\chiz$ and $b\chip$ are forbidden 
so that the decay $\susytop\ra e^+q$ will dominate.

\begin{figure}[htb]
\centerline{\psfig{file=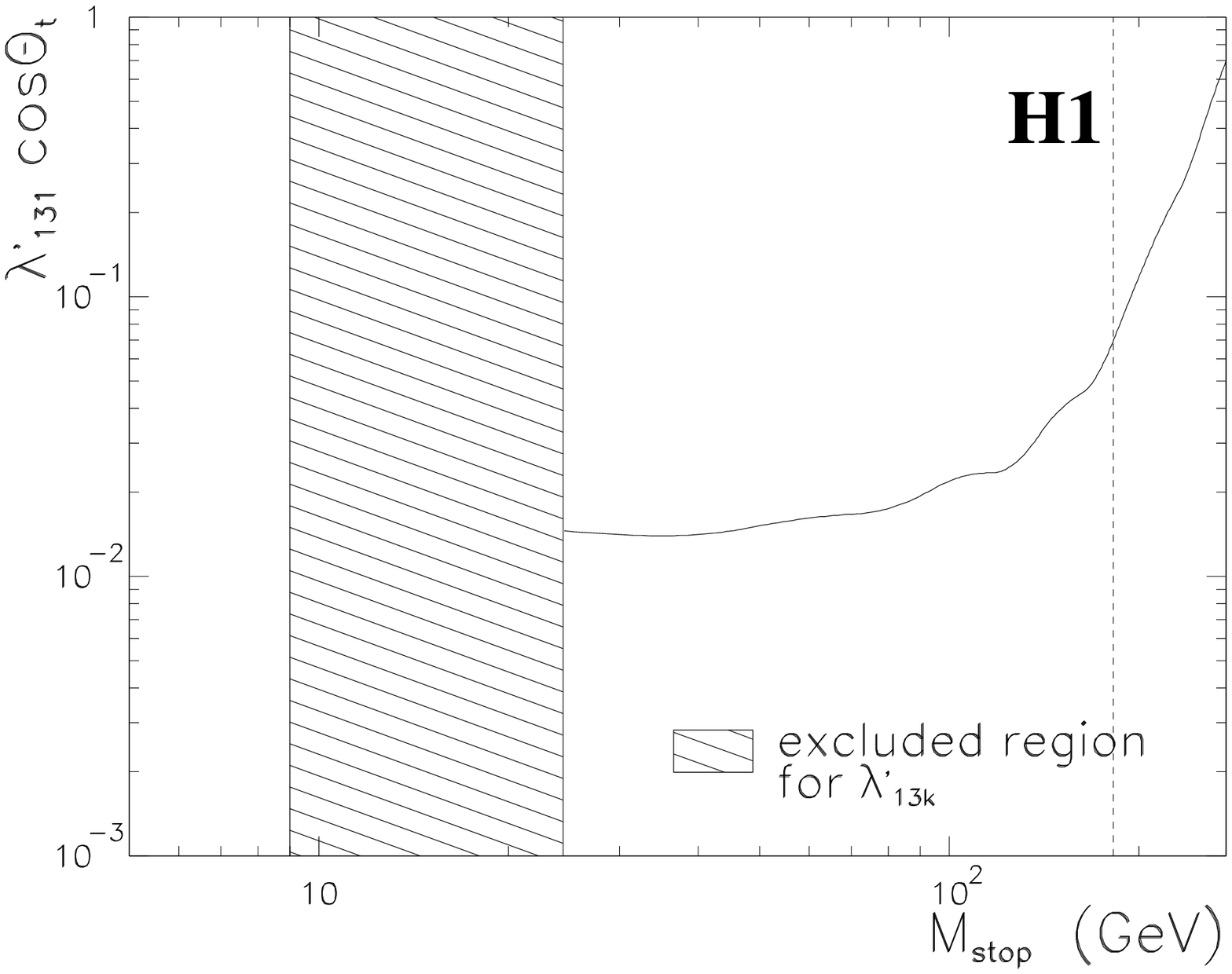,height=9cm}}
\caption{\label{fig:stop_limits}Exclusion limits on the coupling 
$\lambda^\prime_{131}\cos\theta_t$ as a function of the stop mass derived
at 95\% CL. The shaded area is excluded from the pair-production search while
the region above the curve is excluded from single stop production.}
\end{figure}

The final state for such a signature will comprise two isolated and 
acollinear high $E_T$ electrons together with two high $E_T$ jets and no
missing momentum. Details of the analysis are published in~\cite{h1squark}.
As no signal  was observed using \lumi=2.83~\pbarn, exclusion limits were
derived at 95\% CL, combining those channels from the previous 
search in which single \susytop\ is produced. Figure~\ref{fig:stop_limits} 
shows the limits on the coupling
$\lambda^\prime_{131}\cos\theta_t$ as a function of $M_{\susytop}$. The 
exclusion limits from the pair production analysis, indicated by the shaded 
area, are independent of the coupling and exclude masses between 
9 and 24.4 GeV.

The region above the curve in figure~\ref{fig:stop_limits} is excluded by the 
single stop search. For coupling strengths 
$(\lambda^\prime_{131}\cos\theta_t)^2/4\pi
\geq 0.01 \alpha_{em}$ stop masses below 138~GeV are excluded.
In the region where the stop mass limits extend beyond the top mass, the
limits are only valid when $M_{\susytop} < M_{\rm top}+M_{\chiz}$.

\section{Conclusions and future prospects}
The ZEUS and H1 experiments have found no evidence for excited
fermions or supersymmetry using an integrated luminosity of up to 20~\pbarn.
Upper limits at 95\% confidence level have been established on excited 
electron and neutrino production which are more
stringent than the current limits at LEP. The limits on excited quark 
production via electroweak mechanisms complement those measurements made at 
the Tevatron where their strong production is explored. New limits on SUSY 
particle production at HERA have been established within the context of the 
MSSM and R-parity violating  SUSY, including limits on stop production.

The future prospects for discovery at HERA
continue to improve as the luminosity delivered 
by the HERA accelerator continues to increase. In 1997 each experiment 
expects to collect an integrated luminosity 
of $\sim 30~\pbarn$ of $e^+p$ collision data, doubling the total data sample
collected pre-1997. In 1998-99 HERA will operate with an $e^-$ beam and the 
goal is to deliver $\lumi=50~\pbarn$  per experiment, thereby
improving the sensitivity for the search for excited neutrinos, certain
leptoquarks and R-parity violating squarks. 
HERA will undergo a major luminosity upgrade in the shutdown between 
1999-2000, beyond which each experiment expects to collect a total of 
1000~\pbarn.

\end{document}